\documentclass{aa}
\usepackage{graphicx}
\usepackage{natbib}
\newcommand{\um}{\ensuremath{\mu\mbox{m}}}

\begin{document}
   \title{Water ice growth around evolved stars}


   \author{C. Dijkstra
           \inst{1}
           \and
           C.Dominik
           \inst{1}
           \and
           S.N. Hoogzaad
           \inst{1}
           \and
           A. de Koter
           \inst{1}
           \and
           M. Min
           \inst{1}
          }

   \offprints{C. Dijkstra (dijkstra@astro.uva.nl)}

   \institute{Astronomical Institute, ``Anton Pannekoek'', University
              of Amsterdam, Kruislaan 403, NL-1098 SJ Amsterdam, The
              Netherlands
          \and
              Instituut voor Sterrenkunde, Katholieke Universiteit Leuven,
              Celestijnenlaan 200B, B-3001 Heverlee, Belgium
             }

   \date{Received ?? / Accepted ??}

   \abstract{We present a model of the growth of water ice on silicate
     grains in the circumstellar envelopes of Asymptotic Giant Branch
     (AGB) stars and Red Super Giants. We consider the growth of ice
     by gas grain collisions, the thermal evaporation of ice from a
     grain, and sputtering.  Our model contains several improvements
     compared to earlier models, including a detailed treatment of the
     effects of sputtering, a detailed calculation of the radiation
     pressure on the grain, and the treatment of subsonic drift
     velocities. In terms of drift velocity between the grains and gas
     in the envelope, we find that the ice formation process can be
     divided into three regimes: (i) a sputtering dominated regime
     where ice growth is heavily suppressed, (ii) an intermediate
     regime with moderately efficient condensation and (iii) a
     thermally dominated regime where ice formation is highly
     efficient. Sputtering is the critical factor which determines if
     ice formation can occur at all.  We find that in Red Supergiants,
     ice formation is suppressed, while the winds of OH/IR stars allow
     for efficient condensation and can convert significant fractions
     of the available water vapor (tens of percent) into ice mantles
     on grains. Population II AGB stars hardly form ice due to
       their low dust to gas ratios. We also modify an analytical
     equation describing condensation and depletion
     \citep{1985ApJ...292..487J} in order to give reasonable results
     for high and low drift velocities. Initially, ice will condense
     in crystalline form, but continuing condensation at low
     temperatures, and damage caused by interstellar UV photons favor
     the production of amorphous ice as well. We predict that a
     significant fraction of the ice formed will be amorphous.

    \keywords{circumstellar matter -- infrared: stars -- stars: abundances -- stars: AGB and post-AGB -- stars: mass-loss 
            }
   }

   \maketitle
%

\section{Introduction}

\label{sect:introduction}

Asymptotic Giant Branch (AGB) stars represent the final phases of the
evolution of stars with low or intermediate Main Sequence (MS) mass
($1{\ }{\leq}{\ }M<8{\ }M_{\odot}$). During the AGB phase, which lasts
some $10^{6}{\ }\rm{yr}$, stars expel matter at a high rate ($10^{-7}$
to $10^{-4}{\ }\rm{M_{\odot}/yr}$). Mass-loss increases over
time, eventually reaching rates in excess of $10^{-5}{\ }\rm{M_{\odot}/yr}$. 
The outflowing matter creates a dusty molecular circumstellar envelope which 
may completely obscure the central star. As the dust is driven outwards by 
radiation pressure of the central star, it drags the molecules in the envelope
along with it. Given the right conditions, some species of these molecules may
condense onto the grains as ice further out in the outflow. The ice on these 
grains forms an important diagnostics. It can be detected by its infrared 
features and provides information about the physical conditions in the
envelope. Ice mantles may also be chemically important, for example for the 
formation of carbonates \citep{2002Natur.415..295K} or hydrous silicates. At 
the same time, the condensing molecules disappear from the gas, and the 
associated emission of molecular lines becomes weaker. This applies both to
the main ice molecule (e.g. H$_2$O) and derived species (e.g. OH maser 
emission). An understanding of the ice condensation process is therefore 
critical for the interpretation of both the solid-state and the molecular 
diagnostics of AGB envelopes.

Observations reveal that water ice is an important solid state component 
in the circumstellar environments of evolved stars. Water ice is found in the 
environments of AGB stars like OH 26.5+0.6, OH 127.8+0.0, OH 231.8+4.2 
\citep[see][ and references therein]{1990ApJ...355L..27O}, post-AGB stars 
like HD 161796 \citep{2002A&A...389..547H}, planetary nebulae like 
NGC 6302 \citep[see e.g.][]{2001A&A...372..165M}, and at least one Red Super 
Giant, NML Cyg \citep{2002A&A...382..184M}. In these environments it 
has mainly been observed in a crystalline state. This has to a large extend 
been deduced from the presence of both a $43{\ }\rm{{\mu}m}$ and 
$60{\ }\rm{{\mu}m}$ feature in their {\em Infrared Space Observatory} 
\citep[ISO,][]{1996A&A...315L..27K} spectra. Detailed studies of these spectra
can be found in e.g. \citet{B_98_LWS_AGB} (who describes the features for 
different types of sources) and \citet{1999A&A...352..587S} (who describe the 
features for AGB OH/IR stars specifically).

In this paper we present a detailed model which describes the growth
of water ice on silicate grains in the circumstellar envelopes of
oxygen-rich AGB stars. The model is also applicable to Red Super Giants.
The general conditions for the formation of water ice in circumstellar 
envelopes have been described and discussed by \citet{1985ApJ...292..487J}; 
(hereafter refered to as J$\&$M). Their model includes most of the major 
processes involved in the formation of ice. However, as was already pointed 
out by J$\&$M, two ingredients are missing in their model: sputtering (the 
mechanical removal of ice lattice particles from the surface of the grain by 
energetic collisions with gas particles) and subsonic drift velocities (i.e. 
subsonic velocity differences between the dust and gas). We will show, that a 
detailed treatment of these ingredients is necessary if we wish to properly 
describe the ice formation process around evolved stars in a broad set of 
circumstellar environments. Compared to the J$\&$M model, our model also 
improves the calculation of the radiation pressure on the grain, taking into 
account the optical properties of the grain and the spectral energy 
distribution of the driving radiation.

In Sect. \ref{sect:themodel} we give a complete description of the 
model, while in Sect. \ref{sect:ResultsDiscussion} the results of the model 
will be presented and discussed. In Sect. \ref{sect:observations} we apply
our model to an OH/IR star, a Red Super Giant and a population II AGB star and
discuss why ice does or does not grow in the circumstellar environments of 
these stars. The conclusions of this paper are given in Sect. 
\ref{sect:summaryconclusions}. Sect. \ref{sect:ResultsDiscussion} will 
include discussions on 1) the effects of mass-loss rate, grain size and drift
velocity on ice formation, 2) the distance dependence of ice formation and 
the question whether crystalline or amorphous ice will form, and 3) the 
derivation of a generally valid analytical expression for the depletion of 
water vapour from the envelope.

In a future paper (Dijkstra et al., in prep.) we will describe the
observational applications of our model calculations in detail. In that paper
we will use the model to predict the infrared spectra emitted by evolved stars
under the influence of the ice formation process. The resulting spectra will
be compared with observations of individual stars, and used to derive their
stellar parameters.

\section{The model}

\label{sect:themodel}

\subsection{Radiation force, drag force and drift velocity}

\label{sect:FradFdragvdrift}

A silicate grain near an AGB star, will be blown away from the star by 
radiation pressure. As the grain moves away from the star it travels through 
the circumstellar envelope which surrounds the AGB star. On its journey the 
grain will hit gas molecules in the envelope. The collisions create a drag 
force on the grain. We will assume that the drag force will be in equilibrium 
with the radiation force at any time and place. We define $r$ to be the 
distance of the grain from the center of the star.

The radiation force on the grain is given by

\begin{equation}
F_{\rm{rad}}=\frac{4{\pi}^{2}{a}^{2}}{c}\int_{\lambda=0}^\infty Q_{\rm{abs},\lambda}H_{\lambda}\rm{d}\lambda
\label{eq:Frad}
\end{equation}
where $Q_{\rm{abs},\lambda}$ is the absorptivity efficiency of the grain at 
wavelength $\lambda$, and $\rm{H_{\lambda}}$ is the local Eddington flux at
wavelength $\lambda$. For the calculation of
$H_{\lambda}$ we use the radiative transfer programme {\sc modust} 
\citep{2001PhDT..........B}, which models the (emergent) spectrum of a 
spherical dust shell around a central star. $a$ is the radius of the grain and
$c$ is the speed of light. Eq. \ref{eq:Frad} may be simplified to

\begin{equation}
F_{\rm{rad}}=\frac{{Q_{\rm{abs}}}L_{\ast}}{4c}\left(\frac{a}{r}\right)^{2}
\label{eq:Fradsimple}
\end{equation}
where $L_{\ast}$ is the total luminosity of the star, i.e. the luminosity 
integrated over all $\lambda$. $Q_{\rm{abs}}$ is the mean absorptivity 
efficiency integrated over all $\lambda$ with respect to the local energy 
distribution. In Sect. \ref{section:optical} the calculation of 
$Q_{\rm{abs},\lambda}$ will be discussed in detail.

The drag force is given by

\begin{equation}
F_{\rm{drag}}=\pi{a^2}{\rho}v^{2}_{\rm{th}}\sqrt{\frac{64}{9\pi}s^{2}+s^{4}}
\label{eq:Fdrag}
\end{equation}
as discussed by \citet{1992ThesisCarsten}. Here $\rho$ is the mass 
density of the gas in the envelope, while $v_{\rm{th}}$ is its thermal speed.
$s$ is defined to be $v_{\rm{drift}}/v_{\rm{th}}$ where 
$v_{\rm{drift}}$ is the drift
velocity, i.e. the velocity of the dust grain with respect to the gas.

Eq. \ref{eq:Frad} or \ref{eq:Fradsimple} together with Eq. 
\ref{eq:Fdrag} can be solved for the drift velocity of the grain. Setting 
$F_{\rm{rad}}$ equal to $F_{\rm{drag}}$ we obtain

\begin{equation}
v_{\rm{drift}}=v_{\rm{th}}\sqrt{-\frac{32}{9\pi}+\sqrt{\left(\frac{32}{9\pi}\right)^{2}+\left({\frac{F_{\rm{rad}}}{{\pi}a^{2}\rho_{\rm{gas}}{v_{\rm{th}}}^{2}}}\right)^{2}}}.
\label{eq:vdrift}
\end{equation}

$v_{\rm{drift}}$ is an important quantity for the model since, as we will
show later, for supersonic drift velocities it determines both the rate at 
which water molecules are collected from the gas and the rate at which these
molecules are sputtered from the grain's surface again, thus influencing the 
amount of ice that forms. $v_{\rm{drift}}$ also enters the expression for 
$da/dr$, the rate of change in radius of the icemantle as a function of 
distance to the star, which we will derive now.

If $dV$ is the change in volume of the grain as it collects $dN_{\rm{H_{2}O}}$
water molecules, each occupying a volume $V_{\rm{H_{2}O}}$, from the gas in
the envelope, then we have

\begin{equation}
da=\frac{dV}{4{\pi}a^{2}}=\frac{dN_{\rm{H_{2}O}}V_{\rm{H_{2}O}}}{4{\pi}a^{2}}.
\end{equation}
The above equation assumes that all water molecules are distributed
uniformly over the grain surface, i.e. the grain remains spherical at all 
times. For a given time interval $dt$, $dN_{\rm{H_{2}O}}$ is given by (i)
the rate at which water molecules from the gas in the envelope are collected 
by the grain, i.e. the growth rate $Z_{\rm{gr}}$, (ii) the rate at which
water molecules are lost from the grain due to evaporation, i.e. the 
evaporation rate $Z_{\rm{ev}}$, and (iii) the rate at which water molecules 
are lost from the grain due to sputtering, i.e. the sputtering rate 
$Z_{\rm{sp}}$. With the velocity of the dust particle with respect to the star 
given by $v_{\rm{dust}}=dr/dt$ and the density of the ice in the mantle 
given by $\rho_{\rm{ice}}=\frac{18m_{\rm{p}}}{V_{\rm{H_{2}O}}}$, where 
$m_{\rm{p}}$ is the mass of a proton, we find 

\begin{eqnarray}
da&=&\frac{(Z_{\rm{gr}}-Z_{\rm{ev}}-Z_{\rm{sp}})dtV_{\rm{H_{2}O}}}{4{\pi}a^{2}}
\nonumber\\
&=&\frac{(Z_{\rm{gr}}-Z_{\rm{ev}}-Z_{\rm{sp}})V_{\rm{H_{2}O}}}{4{\pi}a^{2}v_{\rm{dust}}}dr,
\end{eqnarray}
yielding

\begin{equation}
\frac{da}{dr}=\frac{9m_{\rm{p}}}{2\pi\rho_{\rm{ice}}v_{\rm{dust}}a^{2}}(Z_{\rm{gr}}-Z_{\rm{ev}}-Z_{\rm{sp}}).
\label{eq:dadr}
\end{equation}

$v_{\rm{drift}}$ enters Eq. \ref{eq:dadr} if we note that it can be 
expressed as $v_{\rm{drift}}=v_{\rm{dust}}-v_{\rm{gas}}$, where 
$v_{\rm{gas}}$ is the speed of the gas in the envelope with respect to the 
star. $v_{\rm{gas}}$ is assumed to be an input parameter of the model. With 
this expression for $\frac{da}{dr}$ we will now study $Z_{\rm{gr}}$, 
$Z_{\rm{ev}}$ and $Z_{\rm{sp}}$ in more detail.

\subsection{The growth rate, $Z_{\rm{gr}}$}

\label{sect:Zgrow}

Assuming a Maxwell velocity distribution for the gas, the rate at which the 
dust grain collects water molecules from the gas is given by 
\citep{1992ThesisCarsten}

\begin{eqnarray}
Z_{\rm{gr}}&=&\frac{{\alpha}{\pi}a^{2}{v_{\rm{th}}^{\rm{H_{2}O}}}}{\sqrt\pi}\frac{n_{\rm{H_{2}O}}}{s_{\rm{H_{2}O}}}\times
\nonumber\\
& &\left[\frac{\sqrt\pi}{2}(2{s^{2}_{\rm{H_{2}O}}}+1)\phi(s_{\rm{H_{2}O}})+s_{\rm{H_{2}O}}{e}^{-{s^{2}_{\rm{H_{2}O}}}}\right]
\label{eq:Zgr}
\end{eqnarray}
where $n_{\rm{H_{2}O}}$ is the number density of water molecules in the gas,
$\alpha$ is the sticking probability of a water molecule on the grain's
surface.

\begin{equation}
s_{\rm{H_{2}O}}=\frac{v_{\rm{drift}}}{v_{\rm{th}}^{\rm{H_{2}O}}},
\end{equation}

\begin{equation}
\phi(s_{\rm{H_{2}O}})=\frac{2}{\sqrt{\pi}}\int_{0}^{s_{\rm{H_{2}O}}} {e}^{-t^{2}}\rm{d}t
\end{equation}
is the error function and 

\begin{equation}
v_{\rm{th}}^{\rm{H_{2}O}}=\sqrt{\frac{2kT_{\rm{gas}}}{18m_{\rm{p}}}}
\end{equation}
is the thermal speed of the water molecules in the gas. $T_{\rm{gas}}$, the 
temperature structure of the gas and dust in the envelope as a function of
distance to the central star, may be given as a power law or calculated using 
radiative transfer codes.

The sticking probability will be assumed in our calculations to
  be equal to one, since most of the ice formation will happen at low
  temperatures and with water molecules sticking to an ice surface.
  Because of the dipole forces in water ice, sticking is likely in
  this case.  However, initially ice will grow on silicate cores and
  possibly on other materials, where sticking has to rely on van der
  Waals forces only.  In this case it is possible that $\alpha$
  will drop below unity.  Also at high drift velocites
  sticking may become inefficient.  While we don't treat the
  velocity dependence of $\alpha$, ice growth becomes automatically
  inefficient at those drift speeds because of sputtering.

For the limiting cases of small and large drift velocities $Z_{\rm{gr}}$ may
be written as

\begin{equation}
Z_{\rm{gr}}=
\left\{
\begin{array}{ll}
{\alpha}4{\pi}a^{2}n_{\rm{H_{2}O}}\frac{v_{\rm{th}}^{\rm{H_{2}O}}}{2\sqrt{\pi}} ~ ~ ~ {\rm :~} s_{\rm{H_{2}O}}\ll1\\
{\alpha}{\pi}a^{2}n_{\rm{H_{2}O}}|v_{\rm{drift}}| ~ ~ ~ {\rm :~} s_{\rm{H_{2}O}}\gg1
\end{array}
\right.
\label{eq:Zgrlimits}
\end{equation}

\subsection{The evaporation rate, $Z_{\rm{ev}}$}

\label{sect:Zevap}

Consider a mass of water ice surrounded by water vapour. The vapour pressure 
of water, ${p_{\rm{vap}}^{\rm{H_{2}O}}}$, is defined to be the equilibrium
pressure at which the rate in which the water ice evaporates from a surface is
equal to the rate in which the water vapour condenses into ice on the surface.
If the water gas pressure is larger (lower) than 
${p_{\rm{vap}}^{\rm{H_{2}O}}}$ ice will grow (evaporate). We can (based on the
Clausius-Clapeyron relation) express ${p_{\rm{vap}}^{\rm{H_{2}O}}}$ as 

\begin{equation}
{p_{\rm{vap}}^{\rm{H_{2}O}}}=p_{\rm{H_{2}O}}\frac{{n_{\rm{vap}}^{\rm{H_{2}O}}}}{n_{\rm{H_{2}O}}}={1.6}\times{10}^{12}{e}^{-(\frac{5600\rm{K}}{T_{\rm{dust}}})}
\label{eq:pH2Ovap}
\end{equation}
where $p_{\rm{H_{2}O}}$ is the water gas pressure in 
$\rm{dyn{\ }{cm}^{-2}}$, ${n_{\rm{vap}}^{\rm{H_{2}O}}}$ is the number 
density of water vapour molecules at 
$p_{\rm{H_{2}O}}={p_{\rm{vap}}^{\rm{H_{2}O}}}$ and $T_{\rm{dust}}$ is the
temperature of the grain \citep{1987Natur.330..550K,2001MasterThesisSeppe}. We
assume that $T_{\rm{dust}}=T_{\rm{gas}}$. We will need 
${p_{\rm{vap}}^{\rm{H_{2}O}}}$ in our calculation of the evaporation rate, 
$Z_{\rm{ev}}$.

The evaporation rate can be calculated from ${p_{\rm{vap}}^{\rm{H_{2}O}}}$,
using that in thermal equilibrium (TE) the growth rate must equal the 
evaporation rate, i.e. $Z_{\rm{gr}}=Z_{\rm{ev}}$. The growth rate for a
grain in TE is given by \citet{1992ThesisCarsten} as (see also Eq. 
\ref{eq:Zgrlimits} in the limit $s_{\rm{H_{2}O}}\ll1$)

\begin{equation}
Z_{\rm{gr}}={\alpha}4\pi{a}^{2}{n_{\rm{vap}}^{\rm{H_{2}O}}}\frac{v_{\rm{th}}^{\rm{H_{2}O}}}{2\sqrt{\pi}}
\label{eq:Zgrsmallvdrift}
\end{equation}

Using $Z_{\rm{ev}}=Z_{\rm{gr}}$, $\alpha=1$ and combining Eqs.
\ref{eq:pH2Ovap} and \ref{eq:Zgrsmallvdrift} we find

\begin{eqnarray}
Z_{\rm{ev}}&=&4\pi{a}^{2}{n_{\rm{vap}}^{\rm{H_{2}O}}}\frac{v_{\rm{th}}^{\rm{H_{2}O}}}{2\sqrt{\pi}} 
\nonumber\\
&=&4\pi{a}^{2}n_{\rm{H_{2}O}}\frac{{p_{\rm{vap}}^{\rm{H_{2}O}}}}{p_{\rm{H_{2}O}}}\frac{v_{\rm{th}}^{\rm{H_{2}O}}}{2\sqrt{\pi}}
\nonumber\\
&=&2\pi{a}^{2}\frac{v_{\rm{th}}^{\rm{H_{2}O}}}{\sqrt{\pi}}\frac{n_{\rm{H_{2}O}}}{S}
\label{eq:Zev}
\end{eqnarray}
where we have defined $S=p_{\rm{H_{2}O}}/{p_{\rm{vap}}^{\rm{H_{2}O}}}$
as the supersaturation ratio. If $S<1$ ice evaporates from the surface, if
$S>1$ it will grow (as discussed at the beginning of this section).

\subsection{The sputtering rate, $Z_{\rm{sp}}$}

As the grain moves outwards it experiences head-on collisions with gas 
particles in the envelope. Most of the gas in the envelope is present in 
molecular form and its main constituents are molecular Hydrogen 
($\rm{H_{2}}$), Helium (He), Carbon monoxide (CO) and molecular Nitrogen 
($\rm{N_{2}}$). As these particles collide with the grain, they may, 
for large enough drift velocities, mechanically remove lattice particles from 
the ice mantle of the grain. This is known as sputtering. We assume that the 
original silicate grain, which is the core of the dust particle, is not 
destroyed by sputtering\footnote{The silicate core is held together by 
chemical bonds, while the ice mantle is held together by physical bonds. 
Chemical bonds are generally stronger than physical bonds. This makes them less
sensitive to sputtering.}. To incorporate the effects of sputtering in our 
model, we follow the treatment by \citet{1993A&A...274..451W}, who describe 
the sputtering process in detail. We have used their equations and will write 
them down below for completeness. We ignore chemisputtering, i.e. the removal
of ice lattice particles by means of chemical reactions with other species.

In order to calculate the sputtering rate for a species $i$, 
$Z_{\rm{sp},\it{i}}$, we first derive the collision rate of the dust grain 
with the species. In the regime of supersonic drift velocities this rate is
given by
                                         
\begin{equation}
Z_{\rm{col},\it{i}}=n_{i}\pi{a}^{2}v_{\rm{drift}}
\end{equation}
where $n_{i}$ is the gas number density of species $i$. The assumption of
supersonic drift velocities at this point does not affect the results of our
model, since in the ice forming regions (as we will show later) sputtering 
only occurs at supersonic drift velocities. Second, we calculate the sputtering
rate for each species using

\begin{equation}
Z_{\rm{sp},\it{i}}=Z_{\rm{col},\it{i}}Y_{\rm{sp},\it{i}}
\end{equation}
where $Y_{\rm{sp},\it{i}}$ is the sputtering yield for species $i$.
$Y_{\rm{sp},\it{i}}$ can be calculated from 
\citep{1978Rothetal,1980JAP....51.2861B,1985AExpr...1..143S}

\begin{equation}
Y_{\rm{sp},\it{i}}=
\left\{
\begin{array}{ll}
0.0064\frac{m_{0}}{m_{\rm{p}}}{{\gamma_{i}}^{5/3}}{(\frac{E_{i}}{E_{\rm{cr},\it{i}}})}^{\frac{1}{4}}{(1-\frac{E_{\rm{cr},\it{i}}}{E_{i}})}^{\frac{7}{2}} 
& \\ $  $ \quad \quad \quad \quad \quad \quad \quad \quad \quad \quad 
{\rm :~} E_{i}>E_{\rm{cr},\it{i}}\\
0 ~ ~ \quad \quad \quad \quad \quad \quad \quad \quad \quad \,
{\rm :~} E_{i}~{\leq}~E_{\rm{cr},\it{i}}
\end{array}
\right.
\label{eq:Ysp}
\end{equation}

In these equations $E_{i}=0.5m_{i}{v_{\rm{drift}}}^{2}$ is the impact 
kinetic energy of the species $i$ as its collides with the grain and 
$\gamma_{i}=\frac{4m_{0}m_{i}}{{(m_{0}+m_{i})}^{2}}$. $m_{i}$ is 
the mass of an individual member of a species $i$, and $m_{0}$ is the 
mass of a (target) lattice particle. In our case this is equal to the mass of 
a water molecule, i.e. $m_{0}=18m_{\rm{p}}$. The sputtering threshold energy, 
$E_{\rm{cr},\it{i}}$, is given by

\begin{equation}
E_{\rm{cr},\it{i}}=
\left\{
\begin{array}{ll}
\frac{E_{\rm{b}}}{\gamma_{i}(1-\gamma_{i})} & {\rm :~} \frac{m_{i}}{m_{0}}{\leq}0.3\\
8E_{\rm{b}}{(\frac{m_{i}}{m_{0}})}^{2/5} & {\rm :~} \frac{m_{i}}{m_{0}}>0.3
\end{array}
\right.
\end{equation}
where $E_{\rm{b}}$ is the surface binding energy of a lattice particle on
the target material. Sputtering shows a threshold behaviour. Only particles 
with sufficient energy, i.e. $E_{i}>E_{\rm{cr},\it{i}}$, are able to remove 
lattice particles. Particles with $E_{i}<E_{\rm{cr},\it{i}}$ are not. The 
value of $E_{\rm{b}}$ differs for each material and is lower for water ice 
($E_{\rm{b}}=0.1{\ }...{\ }0.37{\ }\rm{eV}$) than for silicates like
$\rm{MgSiO_{3}}$ ($E_{\rm{b}}=4.5{\ }\rm{eV}$). Notice that $E_{\rm{b}}$ 
is roughly an order of magnitude larger for $\rm{MgSiO_{3}}$ than water ice, 
which at least qualitatively justifies our assumption above that only the 
icemantle is affected by sputtering and not the silicate core. Representative 
values of $E_{\rm{b}}$ for different species, including those mentioned 
here, are listed in Table 3 of \citet{1993A&A...274..451W}. 

Finally, with the sputtering yield for each species calculated we can now 
calculate the sputtering rate for each individual species and from this the 
total sputtering rate $Z_{\rm{sp}}$ using (if there are $N$ species)

\begin{equation}
Z_{\rm{sp}}=\sum_{i=1}^N{Z_{\rm{sp},\it{i}}}.
\label{eq:Zsp}
\end{equation}

Another process which may lead to the removal of ice from the
  grains is grain-grain collisions.  Grains of different sizes and/or
  optical properties will drift through the enevelope at different
  speeds and can therefore collide at significant speeds.  Since the
  current models assume a single grain size, relative velocities
  between grains are zero and this effect can be ignored.  However, in
  a future study with grain size distributions, it should be taken
  into account.

\subsection{Depletion}

As the grain moves outwards it collects water molecules from the gas, which 
are then no longer available to interact with new grains, i.e. the water 
molecules are depleted from the gas. The effects of depletion are incorporated
into our models, by calculating the effects of the passage of grains on the 
number density of water molecules in the gas, $n_{\rm{H_{2}O}}$. We have 

\begin{equation}
\frac{dn_{\rm{H_{2}O}}}{dr}=-\frac{n_{\rm{dust}}}{v_{\rm{gas}}}\frac{dN_{\rm{H_{2}O}}}{dt}-\frac{2n_{\rm{H_{2}O}}}{r}
\label{eq:depletionequation}
\end{equation}
for the change in $n_{\rm{H_{2}O}}$ as grains move a distance dr away
from the star in a time interval dt and collect $dN_{\rm{H_{2}O}}$
water molecules from the gas. $n_{\rm{dust}}$ is the number density of
dust grains in the envelope.  The first term on the right hand side of
Eq. \ref{eq:depletionequation} takes into account the depletion
effect. The second term takes into account the effect that as grains
move outward, the surrounding gas dillutes into a larger volume and
therefore $n_{\rm{H_{2}O}}$ drops. Dissociation of
  $\rm{H_{2}O}$ molecules by the interstellar UV field may also lead
  to a decrease of the available molecules.  We have ignored this
  effect.  It will only be important for low mass loss rate stars,
  where (as we will show below) sputtering prohibits ice formation.

$n_{\rm{dust}}$ is calculated using 

\begin{equation}
n_{\rm{dust}}=\frac{3\dot{M}f_{\rm{dg}}}{16{\pi}^{2}{a_{0}}^{3}{\rho}_{\rm{sil}}{r}^{2}v_{\rm{dust}}}
\label{eq:ndust}
\end{equation}
where $\dot{M}$ is the gas mass-loss rate, $f_{\rm{dg}}$ is the dust-to-gas ratio,
$a_{0}$ is the radius of the silicate grain, and hence the initial
radius of the core mantle grain that forms as ice grows on it, and 
${\rho}_{\rm{sil}}$ is the bulk density of a silicate grain (which we took
to be $3300{\ }\rm{{kg}{\ }{m^{-3}}}$). Due to the generally small values for 
$f_{\rm{dg}}$ ($f_{\rm{dg}}{\approx}0.01$) we assume that $\dot{M}$ is not just the gas mass-loss 
rate, but the total mass-loss rate.

\subsection{Crystalline or amorphous ice}

\label{sect:crystamorph}

Ice can exist in a crystalline or amorphous state. Crystalline ice can 
form if the mobility of the water molecules on the grain surface is large, and
the flux of water molecules on the surface is relatively low. In this case, 
each water molecule has sufficient time to relax to a low energy configuration
and hence help in the creation of a crystalline lattice. Amorphous ice forms 
when the mobility of the water molecules on the grain surface is small and the
flux of water molecules on the surface is relatively high. Now the water 
molecules have no time to relax to a low energy configuration, resulting in an
amorphous structure. A high (low) mobility of the water molecules can be 
achieved at high (low) temperatures. 

In the infrared, crystalline ice will emit solid state features at 
$43{\ }\rm{{\mu}m}$ and $60{\ }\rm{{\mu}m}$ \citep{1969Bertie}, while 
amorphous ice will emit at $43{\ }\rm{{\mu}m}$ but not at 
$60{\ }\rm{{\mu}m}$ \citep{1999MooreC}. In order to correctly predict and/or 
explain the infrared spectra emitted by the circumstellar envelopes of evolved
stars, and hence the physical and chemical conditions inside these envelopes, 
the question whether crystalline or amorphous ice forms must therefore be 
addressed. We will present the spectra emitted by (the grains in) the 
circumstellar envelopes in a future paper. For completeness however, we 
already include the crystalline/amorphous ice discussion in this paper.

To investigate whether crystalline or amorphous ice will form we check if the 
following condition, originally derived by \citet{1994A&A...290.1009K}, is 
full-filled

\begin{equation}
{\cal{F}}_{\rm{gr}}<{\cal{F}}_{\rm{crit}}
\label{eq:icetypecondition}
\end{equation}
where ${\cal{F}}_{\rm{gr}}$ is the growth flux of water molecules per unit 
area and ${\cal{F}}_{\rm{crit}}$ is the socalled critical flux. 
${\cal{F}}_{\rm{gr}}$ is given by

\begin{equation}
{\cal{F}}_{\rm{gr}}=\frac{Z_{\rm{gr}}}{4\pi{a}^{2}}
\label{eq:condition}
\end{equation}
while ${\cal{F}}_{\rm{crit}}$ is given by

\begin{equation}
{\cal{F}}_{\rm{crit}}=\frac{D_{\rm{s}}(0){e}^{-(\frac{E_{\rm{s}}}{kT_{\rm{dust}}})}}{{a_{\rm{lattice}}}^{4}}
\label{eq:Fcrit}
\end{equation}
where $D_{\rm{s}}(0)=1.74\times{10}^{1}{\ }\rm{m^{2}s^{-1}}$, k is Boltzmann's 
constant and $E_{s}/k=4590{\ }\rm{K}$. $a_{\rm{lattice}}$ is the 
lattice constant of water ice, which is $4.5{\ }\rm{\AA}$. We will once more 
assume that $T_{\rm{dust}}=T_{\rm{gas}}$ (see Sect. \ref{sect:Zevap}). 
Condition \ref{eq:icetypecondition} 
summarizes the above considerations about whether crystalline or amorphous ice
will form. Indeed, if, at a certain temperature (and therefore a certain 
mobility of the water molecules on the surface of the grain), the flux of 
water molecules on the surface is lower (larger) than a certain critical flux,
crystalline (amorphous) ice will form. For completeness, we note that a 
discussion on the formation of crystalline or amorphous Carbon, in essence 
similar to that given here for water ice, is given by 
\citet{1984A&A...132..163G}.

A third mechanism that determines the type of ice that is formed, besides 
fluxes and temperature, is UV radiation from the Interstellar Medium (ISM). 
Experiments done by \citet{1983Lepault} and \citet{1990fspe.rept..193K} 
reveal that UV radiation can alter crystalline ice into amorphous ice when 
the temperature is below $70{\ }\rm{K}$. This happens under the release of 
$\rm{H_{2}}$ gas. Above $70{\ }\rm{K}$ the crystalline water ice is able 
to repair itself from UV radiation damage and remain crystalline. 
\citet{1992ApJ...401..353M} have shown that the unaltered fraction of 
crystalline water ice under the influence of UV radiation, can be given by

\begin{equation}
{\Phi}(D)={e}^{-{\kappa}D}
\label{eq:unaltered}
\end{equation}

In this equation ${\Phi}(D)$ is the unaltered fraction of crystalline water
ice after it has been irradiated with a dose $D$ of UV radiation. $D$ is given
in eV. ${\kappa}$ is given in molecules/eV and is a measure for the amount of 
water molecules in the crystalline lattice that can be reordered into an 
amorphous structure for each eV of UV radiation supplied. ${\kappa}$ depends 
on the temperature of the ice. A graph showing this dependence is given by 
\citet{1992ApJ...401..353M}: ${\kappa}$ increases with decreasing temperature, 
meaning that for lower temperatures one eV of UV radiation can be used to 
amorphousize a larger fraction of ice. 

At present, our model does not include the effects of the UV radiation from 
the ISM, since we expect its effects on the ice formation process to be 
negligible. However, we do recognize its importance regarding the type of ice 
that is created, and hence the spectrum that the final grain will emit. In 
Sect. \ref{sect:ResultsDiscussion} we will therefore include the issue of 
the UV radiation from the ISM into our discussion where needed.

\subsection{Calculation of $Q_{\rm{abs},\lambda}$}

\label{section:optical}

As the grain collects ice it increases in radius and its optical constants, 
and hence $Q_{\rm{abs},\lambda}$, will change. The change in 
$Q_{\rm{abs},\lambda}$ will affect $F_{\rm{rad}}$, and so
$v_{\rm{drift}}$. This in turn will influence the (final) radius of the grain.
We therefore calculate $Q_{\rm{abs},\lambda}$ at each $r$ to incorporate 
this effect. Based on the value of the size parameter, 
$x=\frac{2\pi{a}}{\lambda}$, we either apply the routine {\sc bhcoat} (taken 
from \citet{1983BohrenHuffman} : used when $x<30$) or an effective medium 
calculation in combination with a MIE calculation (\citet{1983BohrenHuffman} :
used when $x{\geq}30$) to determine $Q_{\rm{abs},\lambda}$. The optical 
constants used in our calculations for the silicate core and the ice mantle 
were taken from \citet{1999MNRAS.304..389S} (cold silicate data) and 
\citet{1969Bertie} respectively.

\section{Results and discussion}

\label{sect:ResultsDiscussion}

In this section we present and discuss the results of the ice growth model
described in the previous section. First, we will discuss the effect of 
$\dot{M}$ and the initial grain size, $a_{0}$, on the formation of ice in the 
circumstellar envelope. Second, we will describe the effect of 
$v_{\rm{drift}}$ and $a_{0}$ on the depletion of water vapour from the 
envelope. Third, we discuss the distance dependence of ice formation and 
address the question whether crystalline or amorphous ice will form, followed 
by a discussion on the type of spectra produced by the core mantle grains 
formed in the envelope. Fourth we derive a generally valid analytical 
expression for the depletion of water vapour from the envelope. Finally, we 
apply our model to individual stars and compare our results with observations 
on these stars.

Our discussion on the above topics will largely be based on models of 
circumstellar envelopes with $\dot{M}$ ranging between $10^{-7}$ and 
$10^{-3}{\ }\rm{M_{\odot}/yr}$. The choice of modelling $\dot{M}$ above the 
more conventional upper limit of $10^{-4}{\ }\rm{M_{\odot}/yr}$ is based on 
observations of the post Red Super Giant binary system AFGL 4106 
\citep{1999A&A...350..163M} and the post-AGB stars HD 161796 
\citep{2002A&A...389..547H} and IRAS 16342-3814 (Dijkstra et al., in prep.). 
All these systems are thought to have experienced mass-loss rates of several 
times $10^{-4}{\ }\rm{M_{\odot}/yr}$ up to $10^{-3}{\ }\rm{M_{\odot}/yr}$ at 
some point during their history, or at least have had densities in the wind 
that, assuming spherically symmetric mass-loss, would require mass-loss-rates
of the order of $10^{-3}{\ }\rm{M_{\odot}/yr}$. Calculations of models with 
$\dot{M}$ above $10^{-4}{\ }\rm{M_{\odot}/yr}$ therefore seem justified. The 
models contain a number of assumptions. Each envelope is composed of 
{\em single sized} particles with radii of either $0.01$, $0.10$, $1.00$ or 
$10.0{\ }\rm{{\mu}m}$. The initial abundance of water vapour (by number
with respect to H) is ${\varepsilon}_{0}^{\rm{H2O}}=1\times{10}^{-4}$. The 
dust-to-gas ratio is equal to $f_{\rm{dg}}=0.01$. The velocity of the gas is given by 
$v_{\rm{gas}}=15{\ }\rm{km/s}$. The central star has an effective temperature
of $T_{\rm{eff}}=3500{\ }\rm{K}$ and a radius 
$R_{\star}=300{\ }\rm{R_{\odot}}$, yielding a luminosity 
$L_{\star}{\approx}1.2{\times}10^{4}{\ }\rm{L_{\odot}}$. We will address the 
importance of these assumptions where needed.

We used the radiative transfer programme {\sc modust} (see Sect. 
\ref{sect:FradFdragvdrift}) to provide a correct physical description of the 
temperature and radiative flux in the envelope. For (very) high mass-loss 
rates, and hence high optical depths, the {\sc modust} models do not fully 
converge. When this problem occured, we resolved it by moving the inner radius
of the circumstellar envelope outwards. Our results will not be severely 
affected by this.

\subsection{Mass-loss rate and initial grain size}

\label{sect:MassLossInitialGrainSize}

\begin{figure}
  \caption{The relative increase in the mass of the core mantle grain (when 
           the complete ice mantle has formed) as a function of $\dot{M}$ for 
           initial grain sizes $a_{0}=0.01$, $0.1$, $1.0$ and 
           $10.0{\ }\rm{{\mu}m}$. For more details see Sect. 
           \ref{sect:MassLossInitialGrainSize}.}
  \rotatebox{90}{\resizebox{6.5cm}{!}{\includegraphics{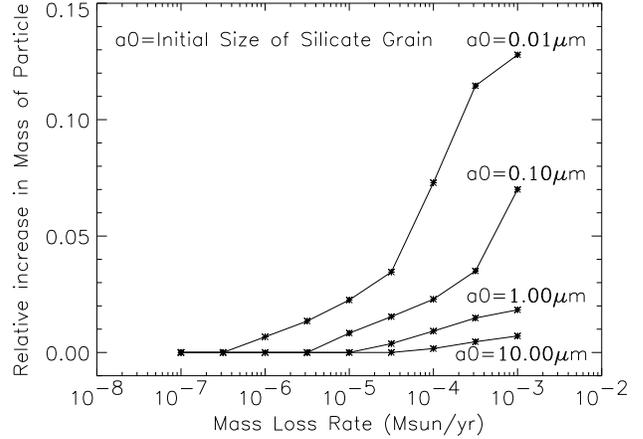}}}
  \label{fig:MdotRelDeltaA}
\end{figure}

In Fig. \ref{fig:MdotRelDeltaA} the relative increase in the mass of the
core mantle grain (when the complete ice mantle has formed), 
$\frac{\Delta{m}}{m}$, is shown as a function of $\dot{M}$ for $a_{0}=0.01$, 
$0.1$, $1.0$, and $10.0{\ }\rm{{\mu}m}$. Fig. \ref{fig:MdotRelDeltaA} shows 
that for a given $a_{0}$ ice formation becomes more efficient as $\dot{M}$ 
increases. For an increasing $\dot{M}$, the density, $\rho$, increases and the
drift velocity, $v_{\rm{drift}}$ decreases\footnote{As $\dot{M}$ and hence 
$\rho$ increaeses the drag force on the grain increases. The drag force must 
balance the radiative force on the grain, which remains approximately 
constant. Therefore $v_{\rm{drift}}$ must decrease.}. The increasing density 
increases the growth rate of the ice on the grain, while the decreasing drift 
velocity lowers the effect of sputtering on the grains, leaving newly formed 
ice layers intact and promoting ice formation as well. Fig. 
\ref{fig:MdotRelDeltaA} also shows that for smaller values of $a_{0}$, 
significant ice formation starts at lower $\dot{M}$. For smaller $a_{0}$, 
$v_{\rm{drift}}$ is lower and sputtering is less important, again leaving 
newly formed ice layers intact. In Sect. \ref{sect:Vdrifta0finfinity} we 
will come back to the importance of sputtering. 

We conclude that small grains have the best ability to grow ice and will most
likely dominate the ice formation process if present in sufficient relative 
numbers. In grain size distributions we generally expect small particles to 
dominate \citep[see e.g.][]{1989A&A...223..227D}. We may thus expect that 
models with small single sized particles and grain size distributions will 
yield similar results. If we for the moment take the behaviour of the smallest
particles as typical, the curve for $a_{0}=0.01{\ }\rm{{\mu}m}$ suggests that 
modest ice growth may already start at mass-loss rates of 
$10^{-6}{\ }\rm{M_{\odot}/yr}$ and that major ice formation only sets in when 
the mass-loss rate is typically of the order of several times $10^{-5}$ to 
$10^{-4}{\ }\rm{M_{\odot}/yr}$, i.e. during the SW phase of an AGB star (see 
Sect. \ref{sect:introduction}). 

\subsection{Mass-loss rate, drift velocity, initial grain size and abundance}

\label{sect:Vdrifta0finfinity}

\begin{figure*}
  \caption{Contours of constant final water vapour abundance relative to the 
           initial abundance, $f(\infty)$, as a function of $a_{0}$ and 
           $\dot{M}$ (upper panel) and $a_{0}$ and $v_{\rm{drift}}$ (lower 
           panel). $f(\infty)$ is given in percentages. Also shown are lines 
           that indicate the critical velocities for sputtering by 
           $\rm{H_{2}}$, He, CO and $\rm{N_{2}}$, and the sound velocity at 
           100 K.}
  \rotatebox{0}{\resizebox{17.5cm}{!}{\includegraphics{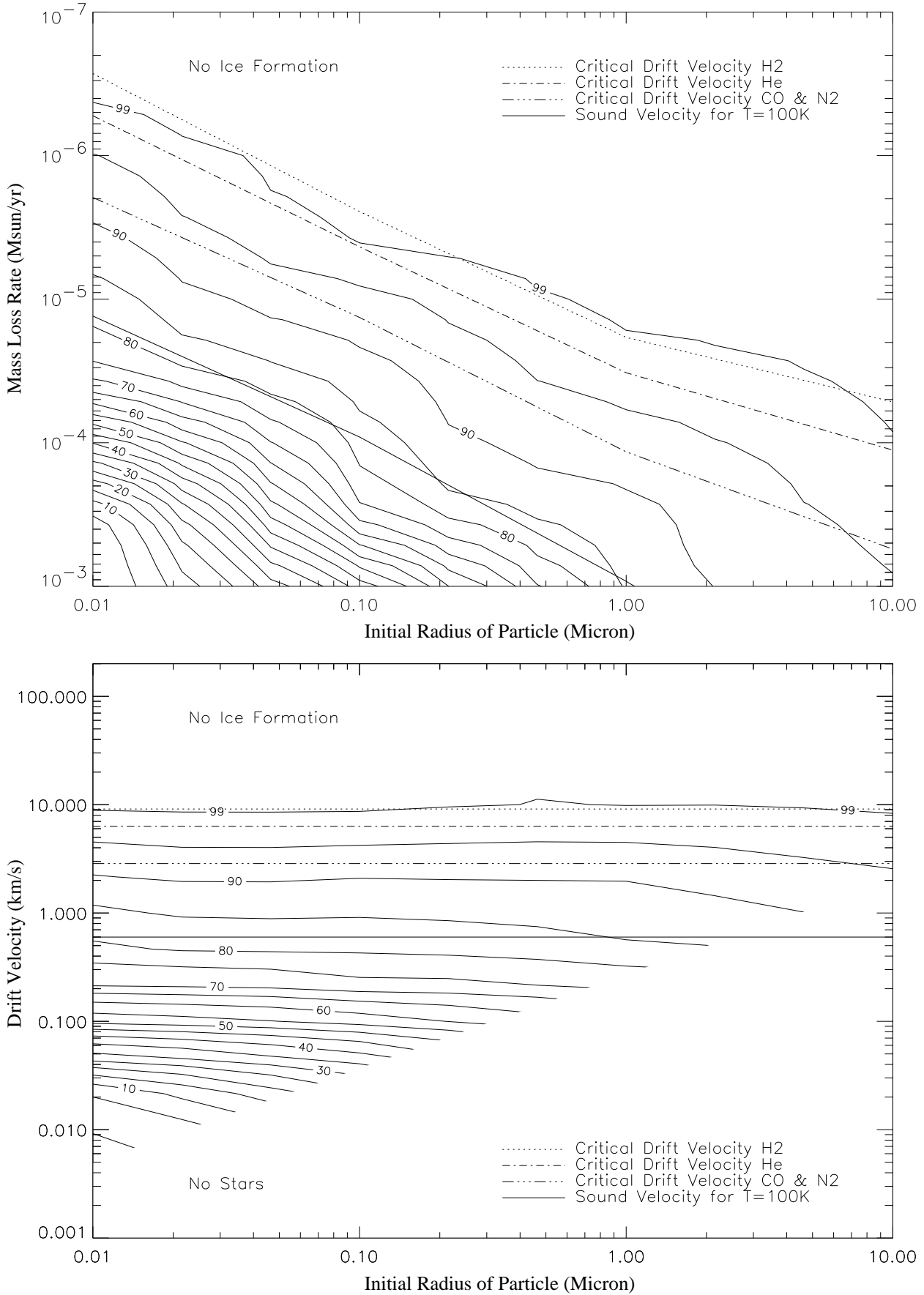}}}
  \label{fig:AfinalVdriftDepletion}
\end{figure*}

Fig. \ref{fig:AfinalVdriftDepletion} shows two contour plots of $f(\infty)$,
the final water vapour abundance in the circumstellar envelope after ice
formation relative to the initial abundance, ${\varepsilon}_{0}^{\rm{H2O}}$. 
In the upper panel of the figure $f(\infty)$ is shown as a function of $a_{0}$
and $\dot{M}$ and given in percentages. The contours, except for the $99\%$ 
contour, are shown at $5\%$ intervals. The unrealistic small scale curvatures 
in some of the contours (see e.g. the $90{\%}$ contour) are due to the 
interpolation of the contour values on the low resolution grid and may 
be ignored. $\dot{M}$ is varied between $10^{-7}{\ }\rm{M_{\odot}/yr}$ and 
$10^{-3}{\ }\rm{M_{\odot}/yr}$. These values correspond to those in Fig. 
\ref{fig:MdotRelDeltaA}. The lower panel shows the same contour plot, but this
time with $v_{\rm{drift}}$ instead of $\dot{M}$ on the horizontal axis. 
$v_{\rm{drift}}$ decreases as $\dot{M}$ increases. Also shown in both contour 
plots are lines indicating where $v_{\rm{drift}}$ reaches the critical 
velocity for sputtering by $\rm{H_{2}}$, $v_{\rm{crit}_{\rm{H_{2}}}}$ 
(${\sim}9{\ }\rm{km/s}$), He, $v_{\rm{crit}_{\rm{He}}}$ 
(${\sim}6{\ }\rm{km/s}$), and CO and $\rm{N_{2}}$, 
$v_{\rm{crit}_{\rm{CO}}}=v_{\rm{crit}_{\rm{N_{2}}}}$ (${\sim}3{\ }\rm{km/s}$),
and the sound velocity at $100{\ }\rm{K}$, $c_{s}$ (${\sim}0.6{\ }\rm{km/s}$).
The 'No Ice Formation' labels in these plots identify the regions where no ice 
formation occurs. The 'No Stars' label in the lower contour plot indicates the
region where $\dot{M}$ exceeds $10^{-3}{\ }\rm{M_{\odot}/yr}$ and no stars are
expected observationally. 

The upper panel of Fig. \ref{fig:AfinalVdriftDepletion} shows that
$f(\infty)$ decreases as $\dot{M}$ increases and/or $a_{0}$ decreases. Also,
ice formation becomes more and more efficient, as is suggested by the
increasing density of the contours. As was discussed in Sect. 
\ref{sect:Vdrifta0finfinity}, an increase in $\dot{M}$ leads to higher 
densities and lower drift velocities, and hence larger growth rates and less 
sputtering. This promotes the formation of ice and makes $f(\infty)$ smaller.
A decrease in $a_{0}$ also leads to lower drift velocities, i.e. less 
sputtering, and (at a fixed $\dot{M}$) it increases the total collecting area 
of the grains. Again, this promotes the formation of ice and makes $f(\infty)$ 
smaller. The obtained values for $f(\infty)$ cover the full range between 
$100$ and $0\%$. $f(\infty)=0\%$, which corresponds to a total freeze out of
all water vapour from the circumstellar envelope, is obtained in the most 
extreme cases, when $\dot{M}{\approx}10^{-3}{\ }\rm{M_{\odot}/yr}$ and 
$a_{0}{\approx}0.01{\ }\rm{{\mu}m}$. 

The contours of $f(\infty)$ in the upper panel of Fig. 
\ref{fig:AfinalVdriftDepletion} run parallel to the lines which indicate where 
the drift velocity reaches the critical velocity for sputtering by 
$\rm{H_{2}}$, He, and CO and $\rm{N_{2}}$, suggesting that the drift velocity
is (indeed) a very important quantity in determining how much ice will grow.
This is illustrated in the lower panel of Fig. 
\ref{fig:AfinalVdriftDepletion}.

In the lower panel the contours depend (almost) only on $v_{\rm{drift}}$ and 
not on $a_{0}$. This allows us to characterize the ice formation process in 
terms of $v_{\rm{drift}}$. The lower panel of Fig. 
\ref{fig:AfinalVdriftDepletion} shows that for any given particle size, ice 
formation is impossible when $v_{\rm{drift}}>v_{\rm{crit}_{\rm{H_{2}}}}$. When
$v_{\rm{drift}}<v_{\rm{crit}_{\rm{H_{2}}}}$ ice formation is switched on, even
though He, CO and $\rm{N_{2}}$ still cause sputtering. The amount of ice
formed in this regime is modest however ($f(\infty)\geq92\%$). Sputtering 
ceases below 
$v_{\rm{drift}}=v_{\rm{crit}_{\rm{CO}}}=v_{\rm{crit}_{\rm{N_{2}}}}$. From now 
on we will define the regime above 
$v_{\rm{drift}}=v_{\rm{crit}_{\rm{CO}}}=v_{\rm{crit}_{\rm{N_{2}}}}$ as the 
sputtering dominated regime. Between $v_{\rm{drift}}{\approx}c_{s}$ and 
$v_{\rm{drift}}=v_{\rm{crit}_{\rm{CO}}}=v_{\rm{crit}_{\rm{N_{2}}}}$
there is an intermediate regime where ice formation becomes increasingly 
more efficient (as can be seen by the increasing contour density), but still 
does not exceed $20{\%}$. At subsonic drift velocities, i.e. when 
$v_{\rm{drift}}<c_{s}$, $f(\infty)$ seems to become very sensitive to 
$v_{\rm{drift}}$, and a total freeze out of water vapour from the envelope 
occurs for the lowest drift velocities 
($v_{\rm{drift}}{\approx}6{\times}{10}^{-3}{\ }\rm{km/s}$). 

We emphasize that in the subsonic or thermally dominated regime, it actually 
is not $v_{\rm{drift}}$ that controls the ice growth, but it is $\dot{M}$ that 
does. Eq. \ref{eq:Zgrlimits} shows that for subsonic drift velocities the
growth rate of ice on a grain does depend on the density $\rho$, and thus 
$\dot{M}$, but not on $v_{\rm{drift}}$. Only the thermal motions 
and density of the gas are important in this regime. The apparent rapid 
decline of $f(\infty)$ with $v_{\rm{drift}}$ is thus simply the effect of the 
increasing mass-loss rate of the star. In the sputtering dominated regime it 
is also $\dot{M}$ that we change, but now it is indeed the effect of 
$v_{\rm{drift}}$ that is truely important, since here it controls the 
sputtering yields (see Eq. \ref{eq:Ysp}), and thus for which gas species 
sputtering is switched on or off. In the intermediate regime, both $\dot{M}$ 
and $v_{\rm{drift}}$ are important. In this regime $\dot{M}$ sets the 
density in the envelope, and thus the growth rate, and it partially sets the 
value for $v_{\rm{drift}}$ ($v_{\rm{drift}}$ depends on more variables than 
$\dot{M}$ alone). $v_{\rm{drift}}$ also sets the growth rate, together with 
the time that the dust spends in a certain (ice formation) region of the 
envelope.

Fig. \ref{fig:AfinalVdriftDepletion} can be used to derive the abundance of 
water vapour in the envelope of an evolved star, keeping in mind the 
assumptions discussed in Sect. \ref{sect:ResultsDiscussion}. The assumptions
that the dust-to-gas ratio is given by $f_{\rm{dg}}=0.01$ and the initial abundance of 
water vapour by ${\varepsilon}_{0}^{\rm{H2O}}=1\times{10}^{-4}$ are an 
important aspect of Fig. \ref{fig:AfinalVdriftDepletion}. When
$f_{\rm{dg}}$ is lower (larger) than $f_{\rm{dg}}=0.01$, the figure
only provides an lower (upper) limit  
on $f(\infty)$, since there will be less (more) dust particles available to 
grow ice on. The effect of a changing ${\varepsilon}_{0}^{\rm{H2O}}$ is not 
trivial since it is dependent on multiple variables that continuously change 
throughout the envelope (see Eq. \ref{eq:depletionequation}). Our models 
suggest that when ${\varepsilon}_{0}^{\rm{H2O}}$ is lower (larger) than 
${\varepsilon}_{0}^{\rm{H2O}}=1\times{10}^{-4}$, Fig. 
\ref{fig:AfinalVdriftDepletion} provides an lower (upper) limit on
$f(\infty)$. This is due to the overall decrease (increase) in the partial 
pressure of the water vapour, which brings the starting point of ice formation
more outwards (inwards) in the envelope, and the decrease (increase) of the 
total collecting area of the grains at any given distance. In Sect. 
\ref{sect:analytical} an analytical equation for $f(\infty)$ will be derived 
that is generally applicable.

\subsection{Crystalline and/or amorphous ice?}

\label{sect:distancecrystamorph}

We have investigated the question whether crystalline or amorphous ice will 
form in the circumstellar envelope. Our model shows that both types of ice may
be present. As a typical example we briefly consider a star with 
$\dot{M}=10^{-5}{\ }\rm{M_{\odot}/yr}$ and $a_{0}=0.01{\ }\rm{{\mu}m}$. We 
find that as the grain moves away from the central star, initially crystalline
ice will be formed, meaning that condition \ref{eq:icetypecondition} is met. 
At a distance of approximately $5100{\ }\rm{AU}$ and onwards, amorphous ice is
formed. Since in condition \ref{eq:icetypecondition}, ${\cal{F}}_{\rm{gr}}$ 
decreases as a function of distance to the star, it must be the decrease in 
$T_{\rm{dust}}$ which causes the breakdown of this condition. In this example, 
$T_{\rm{dust}}\approx65{\ }\rm{K}$ when the transition occurs. So from 
$5100{\ }\rm{AU}$ and onwards, each new water molecule that lands on the 
surface of the grain is immediately immobilised, preventing crystalline ice to
form. We find that (not shown) the grain has reached about $70{\%}$ of its 
final radius, at the time the transition occurs. This means that a substantial
fraction of the grain will be composed of amorphous ice, in this case a volume
fraction of $100(1-{(70/100)}^{3})\approx66{\%}$.

\begin{figure*}
  \caption{Crystalline and amorphous ice formation, for stellar
      envelopes with 0.01\um (left panel) or 0.1\um (right panel)
      grains. Wind trajectories run as vertical lines through this
      diagram from bottom to top.  The different lines show locations where
      conditions for ice formation change (see text).  The shaded area
      indicates where crystalline ice is formed and can survive.}
\vspace{0.2cm}\rotatebox{90}{\resizebox{13.75cm}{!}{\includegraphics{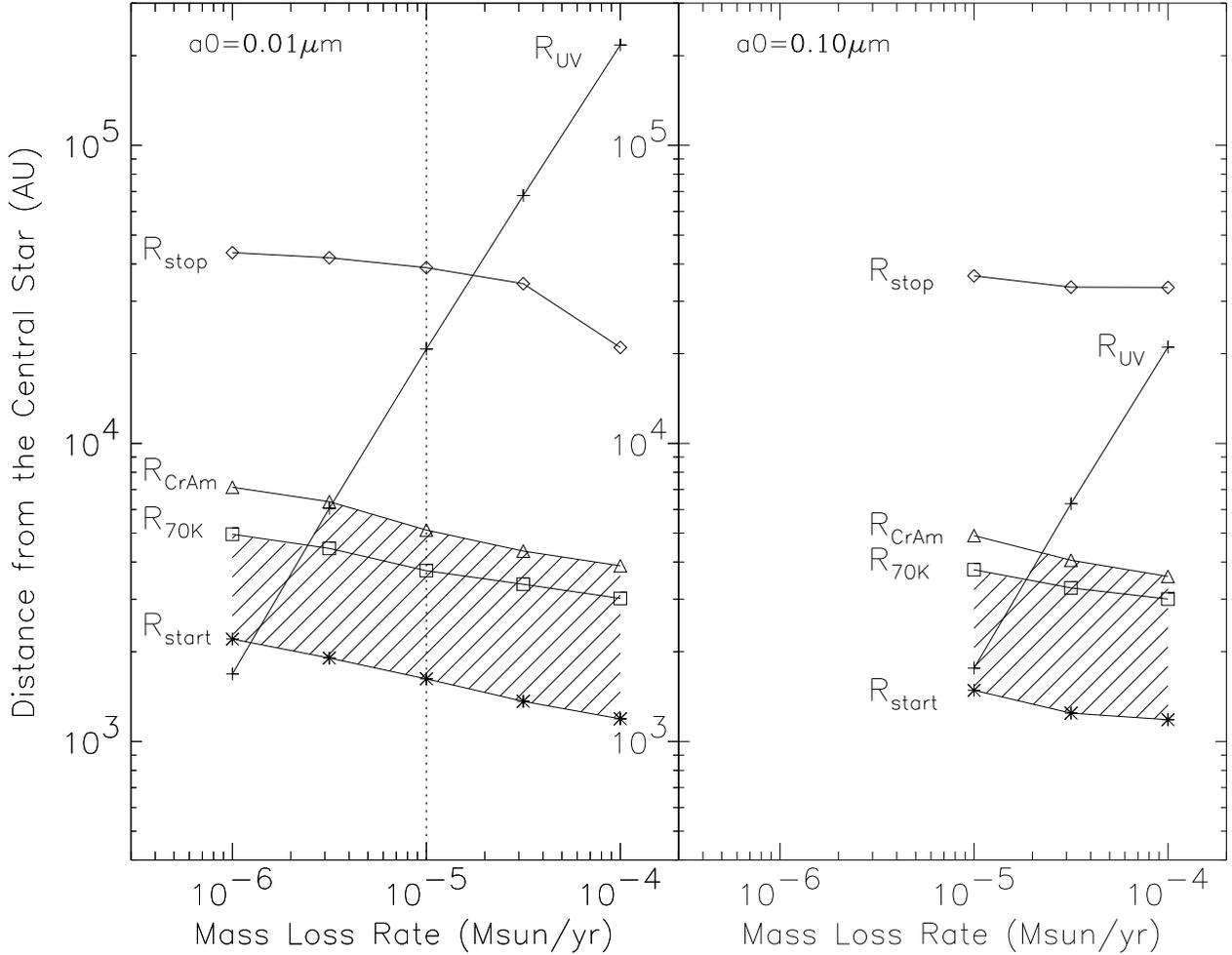}}}
  \label{fig:Distances}
\end{figure*}

Fig. \ref{fig:Distances} shows the dependence of ice formation on the
distance towards the central star. For a given initial grain size, the figure
shows, as a function of mass-loss rate:

\begin{itemize}
\item [.] $R_{\rm{start}}$, the distance where ice formation starts (star 
symbols) 
\item [.] $R_{\rm{stop}}$, the distance where ice condensation
  terminates because of too low H$_2$O densities (diamond symbols).
\item [.] $R_{\rm{70K}}$, where the temperature drops below 70K.  At
  higher temperatures, UV damage in crystalline ice is immediately
  repaired, while at lower temperatures, UV radiation converts
  crystalline ice into amourphous ice (squares).
\item [.] $R_{\rm{CrAm}}$, the distance outside which ice already
  condenses in amorphous form (triangles).
\item [.] $R_{\rm{UV}}$, the distance to which interstellar UV
  radiation can penetrate (crosses).  Outside this radius, ice will be
  amorphisized unless the temperature is above 70K. 
\end{itemize}
We have defined $R_{\rm{stop}}$ to be the distance where the grain has reached
$98{\%}$ of its final size. We determine the distance to which the UV 
radiation can penetrate into the envelope, $R_{\rm{UV}}$, by calculating where
the UV optical depth of the envelope, ${\tau}_{\rm{UV}}$ (at wavelengths 
$\lambda\leq3900\rm{\AA}$), reaches unity. Generally, at a distance $r$ from 
the central star we have (using Eq. \ref{eq:ndust})

\begin{equation}
\tau_{\rm{UV}}(r){\equiv}\int_{r}^\infty Q_{\rm{UV}}{\pi}{a^{2}_{0}}n_{\rm{dust}}(r')dr'=\frac{3Q_{\rm{UV}}f_{\rm{dg}}\dot{M}}{16{\pi}a_{0}{\rho}_{\rm{sil}}v_{\rm{dust}}r}
\label{eq:tauISM}
\end{equation}
and thus 

\begin{equation}
r=\frac{3Q_{\rm{UV}}f_{\rm{dg}}\dot{M}}{16{\pi}a_{0}{\rho}_{\rm{sil}}v_{\rm{dust}}{\tau}_{\rm{UV}}(r)}.
\label{eq:startISM}
\end{equation}
From this we calculate $R_{\rm{UV}}{\equiv}r({\tau}_{\rm{UV}}=1)$. We assume 
that $Q_{\rm{UV}}$, the absorptivity efficiency for UV radiation by the grain,
is given by $Q_{\rm{UV}}=1$ and $v_{\rm{gas}}=15{\ }\rm{km/s}$. The shaded 
areas in Fig. \ref{fig:Distances} indicate those regions where crystalline
ice is formed. In the unshaded regions amorphous ice forms.

Fig. \ref{fig:Distances} shows that increasing the mass-loss rate generally
moves $R_{\rm{start}}$, $R_{\rm{stop}}$, $R_{\rm{70K}}$ and 
$R_{\rm{CrAm}}$ inwards. So, ice formation starts closer to the central
star, the final particle size will be reached earlier, and the transition
point where amorphous ice instead of crystalline ice will be formed moves 
inwards. Meanwhile, $R_{\rm{UV}}$ moves outwards, due to the increasing
optical depth of the circumstellar envelope. For envelopes with low optical 
depths (i.e. envelopes with low mass-loss rates and or large dust particles), 
the effect of the UV radiation is generally important throughout most of the 
envelope, and will most likely contribute to the formation of amorphous ice. 
However, the effect of the UV radiation can be reversed in regions
close to the central star, where the temperature exceeds $70{\ }\rm{K}$.

An increase of the initial grain size generally has a modest effect on the
distance dependence of ice formation. Comparing the two panels in Fig. 
\ref{fig:Distances}, it can be seen that $R_{\rm{start}}$, $R_{\rm{stop}}$, 
$R_{\rm{70\rm{K}}}$ and $R_{\rm{CrAm}}$ only show minor changes in general. 
However, $R_{\rm{UV}}$ changes over an order of magnitude, getting closer to 
the star as $a_{0}$ increases. This is again due to the decrease in the 
optical depth of the envelope. The effect of UV radiation from the ISM on the 
type of ice that forms is thus most important for stars with on average large 
particles in their envelopes.

As an AGB star evolves its mass-loss rate increases over time. Fig. 
\ref{fig:Distances} then suggests that the nature of the ice formation process
will change over time as well. So, to get a qualitative impression on the 
evolution of the distance dependence of ice formation, one might wish to 
substitute the mass-loss rate on the horizonatal axis by a measure for the 
time that the AGB stars has spent on the AGB. It must be noted however, that 
as the star evolves, its luminosity will increase as well, an effect that is 
not incorporated in Fig. \ref{fig:Distances}.

To illustrate the use of Fig. \ref{fig:Distances}, we will look at an example
given by the dotted line in the left panel. This example shows how the 
distance dependence of the ice formation process can be studied for a star 
which loses mass at a rate of $\dot{M}=10^{-5}{\ }\rm{M_{\odot}/yr}$ and has 
particles of radius $a_{0}=0.01{\ }\rm{{\mu}m}$ in its envelope. As a grain 
moves away from this star, the first ice will condense on its surface at 
${\sim}1500{\ }\rm{AU}$. This ice will be crystalline. In this model we found
that the temperature at the start of ice formation is approximately 94 K (not 
shown). For the other models we typically we found values between 92 and 
104 K. At ${\sim}5000{\ }\rm{AU}$ the temperature becomes too low to form 
crystalline ice. Instead amorphous ice will form. Finally, from 
${\sim}20000{\ }\rm{AU}$ onwards, interstellar UV photons will (start to) 
bombard the surface of the grain, most likely destroying all of the 
crystalline ice that is present. At ${\sim}40000{\ }\rm{AU}$ most of the ice 
will have formed on the surface. Notice that if we had taken a star with 
$\dot{M}=10^{-4}{\ }\rm{M_{\odot}/yr}$, the UV photons of the ISM would only 
have become important after ice formation had already ceased. Again, any 
crystalline ice still present in the grain would most likely be destroyed.

\subsection{A recipe for making ice}

\label{sect:analytical}

J$\&$M derived an equation for the final abundance, $f(\infty)$ (also see 
Sect. \ref{sect:Vdrifta0finfinity}), of a condensable species, like water, in
the outflow of an evolved star. The model of J$\&$M considers the outer 
circumstellar envelope of the star, where the matererial has reached its 
terminal outflow velocity, $v_{\rm{gas}}$. It is assumed that there is a 
condensation radius, $r_{0}$, such that for $r<r_{0}$, there is no 
condensation, while for $r>r_{0}$ almost every molecule sticks to the grain. 
Ice is collected by the grains through gas grain collisions and it is removed 
by thermal evaporation. The depletion of water from the gas phase is taken 
into account as well. Two very important assumptions of the J$\&$M model are 
that 1) sputtering is unimportant, since J$\&$M expect the drift velocities of 
the grains to be less than $5{\ }\rm{km/s}$, and that 2) the grains are moving 
through the gas at supersonic drift velocities. Based on our results in Sect.
\ref{sect:Vdrifta0finfinity} we can predict that assumptions 1) and 2) will 
cause the J$\&$M model to break down in respectively the sputtering dominated 
and thermally dominated regime. In the sputtering dominated regime, J$\&$M 
will predict ice growth, while sputtering will heavily supresses or completely 
block ice growth. In the thermally dominated regime the growth rate of the
ice (also see Eq. \ref{eq:Zgrlimits}) will be underestimated, and hence 
the amount of ice that grows. The equation of J$\&$M will only be valid in
the intermediate regime, where assumptions 1) and 2) do apply. In this 
section we will study the equation given by J$\&$M, by applying it to
the circumstellar envelopes described at the beginning of Sect. 
\ref{sect:ResultsDiscussion}. We will compare its predicted depletion for these
envelopes to the depletion predicted by our model calculations of these 
envelopes. Based on this comparison, we will provide a modified version of 
the J$\&$M equation that is valid in all regimes, i.e. the sputtering 
dominated, intermediate and thermally dominated regime.

According to J$\&$M the fraction of a species that remains in the gas phase, 
$f(\infty)$, is given by

\begin{equation}
f(\infty)=\exp(-{\alpha}(v_{\rm{drift}}/v_{\rm{gas}}){\tau}_{\rm{UV}}(r_{0})/Q_{\rm{UV}})
\label{eq:finfinity}
\end{equation}
where $Q_{\rm{UV}}\approx1$ and ${\tau}_{\rm{UV}}(r_{0})$ is a measure of the 
UV optical depth between $r_{0}$ and $\infty$. All other symbols have their 
usual meaning. The depletion of the species from the gas phase is (in general)
given by

\begin{equation}
\Delta=1-f(\infty)
\label{eq:depletion}
\end{equation}

\begin{figure*}
  \caption{(Left panel) The depletion (in percentages), as originally 
           predicted by J$\&$M, $\Delta_{\rm{JM}}$, against the depletion 
           calculated by our model, $\Delta$. The solid line shows where 
           $\Delta_{\rm{JM}}=\Delta$. The inset shows a close up of the 
           plot near the origin. (Right panel) Same as the left panel, but 
           this time the constraints from Eq. \ref{eq:constraints} are 
           forced upon Eq. \ref{eq:finfinity} (or better, Eq. 
           \ref{eq:finfinityadjusted}). For details and a dicussion see 
           Sect. \ref{sect:analytical}.}
  \hspace{0.4cm}\rotatebox{90}{\resizebox{10cm}{!}{\includegraphics{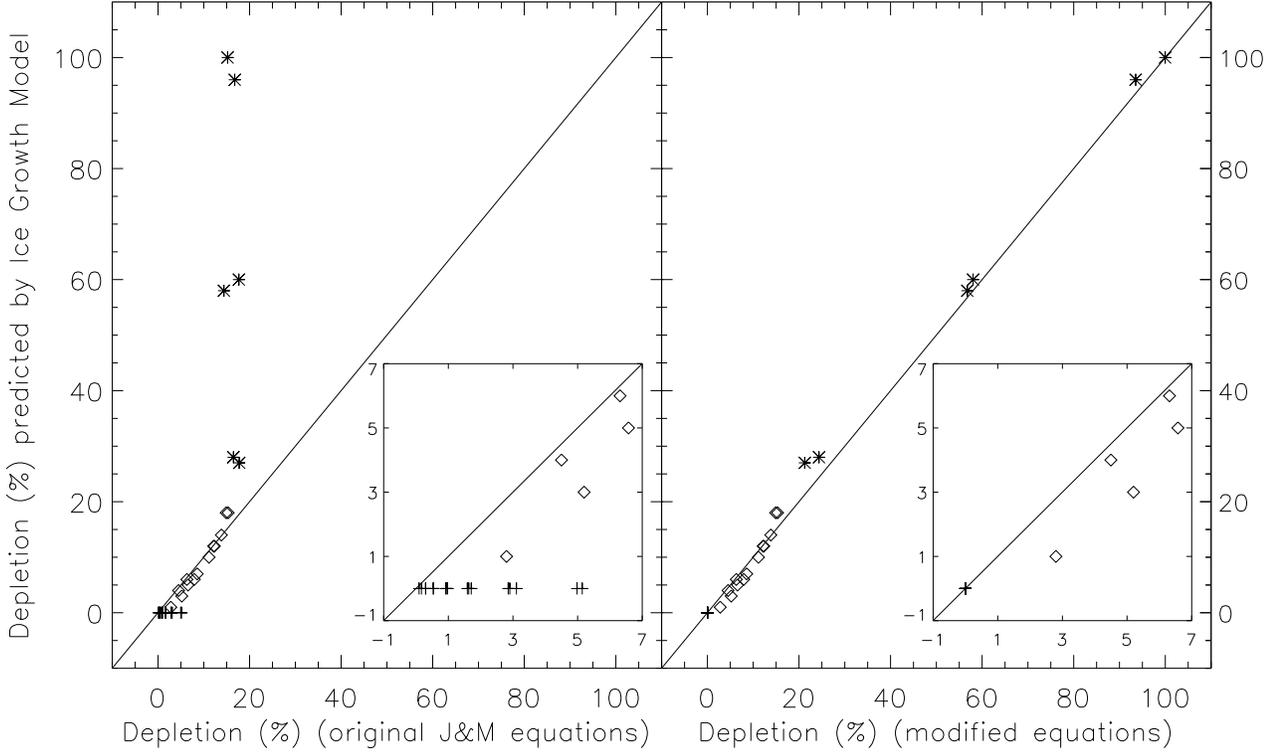}}}
  \label{fig:Analytical}
\end{figure*}

In the left panel of Fig. \ref{fig:Analytical} we plot the depletion as 
predicted by J$\&$M, $\Delta_{\rm{JM}}$, against the depletion predicted by 
our model, $\Delta$, for the circumstellar envelopes described at the 
beginning of Sect. \ref{sect:ResultsDiscussion}. $\Delta_{\rm{JM}}$ (plotted 
on the horizontal axis) and 
$\Delta$ (plotted on the vertical axis) are both given in percentages. The 
solid line shows where $\Delta=\Delta_{\rm{JM}}$. The inset shows a close 
up view of the plot near the origin. It can be seen that for 
$\Delta>20{\%}$ the J$\&$M model underestimates the depletion, i.e. it 
overestimates the true $f(\infty)$. The points for which this happens are 
indicated with stars. Also, for $\Delta=0$, the J$\&$M model predicts that the
depletion is larger than zero, i.e. $\Delta_{\rm{JM}}>0$. Here J$\&$M thus 
overestimate $\Delta$ and underestimate the true $f(\infty)$. The points for 
which this happens are indicated with crosses. Finally, for 
$0<\Delta\leq20{\%}$, the predictions of the J$\&$M model generally agree with
our calculations (these points are indicated with diamonds). For small/large 
values of $\Delta$ in this range, J$\&$M predict somewhat larger/lower 
values for the depletion however. 

For reasons discussed at the begining of this section, the disagreement 
between the J$\&$M model and our model indeed occurs in the sputtering 
dominated regime (crosses) and the thermally dominated regime (stars), where we
respectively have small and large depletions (also see Fig. 
\ref{fig:AfinalVdriftDepletion} in Sect. \ref{sect:Vdrifta0finfinity}). The
intermediate regime (diamonds), where we have moderate depletions, is 
reasonable well described.

The discrepancy in the thermally and sputtering dominated regimes can be fixed 
once it is realized that $v_{\rm{drift}}$ in Eq. \ref{eq:finfinity} is in 
fact the collecting speed of water molecules on the grain, $v_{\rm{collect}}$. 
In the thermally dominated or subsonic regime this collecting speed is set by 
the thermal motions of the gas. The grain is practically at rest with respect 
to the gas and only the thermal motions of the gas particles determine the rate
at which ice is formed (this is expressed in Eq. \ref{eq:Zgrlimits}; see 
the case where $s_{\rm{H_{2}O}}\ll1$ in this equation). In the sputtering 
dominated regime this collecting speed may effectively be set to zero since 
there is no net ice formation due to sputtering (i.e. no net collection of 
water molecules on the surface). In the intermediate regime the collection
speed is simply the drift velocity of the grain. We thus rewrite Eq. 
\ref{eq:finfinity} into 

\begin{equation}
f(\infty)=\exp(-{\alpha}(v_{\rm{collect}}/v_{\rm{gas}}){\tau}_{\rm{UV}}(r_{0})/Q_{\rm{UV}})
\label{eq:finfinityadjusted}
\end{equation}
with

\begin{equation}
v_{\rm{collect}}=
\left\{
\begin{array}{ll}
\max(v_{\rm{drift}},v_{\rm{th}}^{\rm{H_{2}O}}) & \quad \quad : v_{\rm{collect}}<v_{\rm{crit}_{\rm{H_{2}}}}\\
0 & \quad \quad : v_{\rm{collect}}>v_{\rm{crit}_{\rm{H_{2}}}}.\end{array}
\right.
\label{eq:constraints}
\end{equation}
Here $v_{\rm{crit}_{\rm{H_{2}}}}$ is ${\sim}9{\ }\rm{km/s}$ (see Sect. 
\ref{sect:Vdrifta0finfinity}) and we took 
$v_{\rm{th}}^{\rm{H_{2}O}}\approx0.30{\ }\rm{km/s}$. With these new equations 
the predictions by J$\&$M are in better agreement with our model calculations.
This is illustrated in the right panel of Fig. 
\ref{fig:AfinalVdriftDepletion} where we have corrected for the effects of 
subsonic drift and sputtering. Notice that the thermally dominated regime and 
the sputtering dominated regime are much better described now. 

In a part of the sputtering dominated regime, the predictions of J$\&$M still 
suffer from sputtering by He, CO and $\rm{N_{2}}$ at small values of $\Delta$. 
In the intermediate regime they suffer from the onset of the thermal regime
at the largest values of $\Delta$. This leads to respectively somewhat 
larger and smaller values predicted by J$\&$M compared to our 
models. The differences are very modest however.

In order to calculate $v_{\rm{drift}}$ in Eq. \ref{eq:constraints} for 
the intermediate regime, we may use

\begin{equation}
v_{\rm{drift}}=\sqrt{\frac{Q_{\rm{abs}}L_{\ast}v_{\rm{gas}}}{\dot{M}c}}.
\label{eq:vdriftintermediateregime}
\end{equation}
This equation is an approximation of Eq. \ref{eq:vdrift}, valid when 
$v_{\rm{drift}}>v_{\rm{th}}$. For a discussion on this equation and a 
derivation of it from first principles see for example 
\citet{1977ApJ...215..781K} and \citet{1985ApJ...292..487J}.

In summary we conclude that in order to calculate $f(\infty)$ for a given 
(AGB) star, the original equation for $f(\infty)$ as derived by J$\&$M will 
not suffice in many cases, unless thermal motions and sputtering are taken 
into account. This can be done by using the modifications given in Eqs. 
\ref{eq:finfinityadjusted} and \ref{eq:constraints}. In the next section we 
will apply the results of our model calculations, and the modified equation 
for $f(\infty)$ derived in this section, to individual stars.

\section{The model compared with observations : \\ modeling individual stars}

\label{sect:observations}

In this section we will calculate the depletion of water ice from the 
circumstellar envelopes of different types of stars, based on Eqs. 
\ref{eq:tauISM} and \ref{eq:finfinityadjusted} thru
\ref{eq:vdriftintermediateregime}. We will look at some of the details of the
ice formation process around the modeled stars and discuss why ice does or 
does not grow. As examples we will consider an OH/IR star, a Red Super Giant 
(RSG) and a population II AGB star.

The depletion, $\Delta$, will be calculated in the following way. First, we 
determine $v_{\rm{drift}}$ from Eq. \ref{eq:vdriftintermediateregime} 
and we check condition \ref{eq:constraints} to obtain the correct 
$v_{\rm{collect}}$. Second, we calculate ${\tau}_{\rm{UV}}(r_{0})$ using 
Eq. \ref{eq:tauISM}, with $v_{\rm{dust}}=v_{\rm{drift}}+v_{\rm{gas}}$. 
Third, we use $v_{\rm{collect}}$ and ${\tau}_{\rm{UV}}(r_{0})$ to derive 
$f(\infty)$ in Eq. \ref{eq:finfinityadjusted}. Finally, we use Eq. 
\ref{eq:depletion} to find $\Delta$. All our models in this section will 
assume that $a_{0}=0.1{\ }\mu$m, 
${\rho}_{\rm{sil}}=3300{\ }\rm{{kg}{\ }{m^{-3}}}$, $Q_{\rm{UV}}=1$, 
$Q_{\rm{abs}}=0.02$ (here we follow J$\&$M) , $\alpha=1$ and $f_{\rm{dg}}=0.01$. 

For the condensation radius we take $\rm{r_{0}=1600{\ }\rm{AU}}$ in the case 
of the OH/IR and population II AGB star. This choice is based on the results 
of Fig. \ref{fig:Distances} and seems to be a reasonable value when compared
with observations (see e.g. \citet{1985ApJ...294..242S}). In the case of a 
RSG we will take $\rm{r_{0}=16000{\ }\rm{AU}}$ (see e.g. J$\&$M and references 
therein).

\subsection{A typical OH/IR star}

\label{sect:OH/IRstar}

A typical OH/IR star has a luminosity of $L_{\ast}=10^{4}{\ }L_{\odot}$. It 
loses mass with rates between $\dot{M}=10^{-5}\rm{M_{\odot}/yr}$ and 
$\dot{M}=10^{-4}\rm{M_{\odot}/yr}$. The outflow speed of the gas is typically 
$v_{\rm{gas}}=15{\ }\rm{km/s}$. Taking $r_{0}{\approx}1600{\ }\rm{AU}$ we find
that the depletion of water vapour is given by $\Delta{\approx}5{\%}$ and 
$\Delta{\approx}15{\%}$ for $\dot{M}=10^{-5}\rm{M_{\odot}/yr}$ and 
$\dot{M}=10^{-4}\rm{M_{\odot}/yr}$ respectively.

We thus predict that a typical OH/IR star may show depletions of water vapour 
in its envelope up to at least $15{\%}$. In the above calculations we find 
that the drift velocity is in the order of 
$v_{\rm{drift}}{\approx}3{\ }\rm{km/s}$ or lower. This is near or in the 
intermediate regime where sputtering is unimportant and the grain moves 
supersonically through the gas. The absence of sputtering allows the ice 
formation to occur. The predicted presence of ice is in qualitative agreement
with the infrared spectra of many OH/IR stars, which indeed reveal the 
spectral bands of water ice with a variety of band strengths (see e.g.
\citet{1999A&A...352..587S}). Examples of typical OH/IR stars modeled in this 
section are AFGL 5379 \citep[e.g.][]{2001MNRAS.326..490O} and OH 26.5+0.6 
\citep[e.g.][]{1998AJ....115.2509M}.

\subsection{A typical Red Super Giant}

\label{sect:RSG}

A RSG in some sense resembles the massive counterparts of AGB stars. Their MS 
mass is in the range $8{\ }M_{\odot}{\leq}M<60{\ }M_{\odot}$), and like AGB 
stars they are surrounded by a dusty envelope. It is interesting to see if ice
can grow near these stars.

A typical RSG star has a luminosity of $L_{\ast}=2{\times}10^{5}{\ }L{\odot}$.
It may lose mass with typical rates between $\dot{M}=10^{-5}\rm{M_{\odot}/yr}$
and $\dot{M}=10^{-4}\rm{M_{\odot}/yr}$. In extreme cases the mass-loss rate
may near values of $\dot{M}=10^{-3}\rm{M_{\odot}/yr}$ (see e.g. AFGL 4106 
\citep{1999A&A...350..163M}). The outflow speed of the gas is typically 
$v_{\rm{gas}}=30{\ }\rm{km/s}$. Taking $r_{0}{\approx}16000{\ }\rm{AU}$ we 
find that the depletion of water vapour is given by $\Delta=0{\%}$, 
$\Delta{\approx}2{\%}$ and $\Delta{\approx}8{\%}$ for 
$\dot{M}=10^{-5}\rm{M_{\odot}/yr}$, $\dot{M}=10^{-4}\rm{M_{\odot}/yr}$ and 
$\dot{M}=10^{-3}\rm{M_{\odot}/yr}$ respectively.

Based on the results above we predict that near RSGs hardly any water ice
will grow. Typically we may expect $\Delta{\leq}2{\%}$. Only in extreme cases
up to about $8{\%}$ of the water vapout may be depleted into ice. The main 
reason for the inability of RSGs to form ice is that the drift velocity of the
dust is generally too large. The above calculations show that for 
$\dot{M}{\leq}10^{-4}\rm{M_{\odot}/yr}$ the drift velocity will typically be
$v_{\rm{drift}}{\approx}5{\ }\rm{km/s}$ or higher. This is in the sputtering
dominated regime where sputtering heavily suppresses or completely blocks
ice growth. The large drift velocities are mostly due to the high luminosities
of RSGs. Only for extreme mass-loss rates, like 
$\dot{M}=10^{-3}\rm{M_{\odot}/yr}$, efficient ice formation is possible since 
$v_{\rm{drift}}{\approx}1.5{\ }\rm{km/s}$ (intermediate regime) and hence 
sputtering is once more unimportant. 

The presence of water ice towards typical RSGs (e.g. $\alpha{\rm{\ }Ori}$ and 
VY CMa) has to our current knowledge not been frequently reported in 
literature, although in the case of NML Cyg water ice has been detected by
\citet{2002A&A...382..184M}. Post-RSGs with signs that they had high mass-loss 
rates when they where still in the RSG phase do indeed reveal the presence of 
ice (e.g. AFGL 4106, HD 179821 and IRC+10420 \citep{2002A&A...382..184M}). 
This is in agreement with our findings in this section. 

\subsection{A typical population II AGB star}

\label{sect:popIIAGB}

Finally we consider population II AGB stars which generally
  have low metallicities.  The ice formation in this case is limited
  by the amount of dust present in the outflow.  The dust-to-gas ratio
  is at most equal to the metalicity, which can be as low as
  $Z=1\times10^{-4}$.  Taking $f_{\rm{dg}}=10^{-4}$, 
  $r_{0}{\approx}1600{\ }\rm{AU}$, $v_{\rm{gas}}=15{\ }\rm{km/s}$,
  $L_{\ast}=4{\times}10^{3}\,L_{\odot}$, and
  $\dot{M}=10^{-4}\rm{M_{\odot}/yr}$ we find that the drift velocity
  of the dust is $v_{\rm{drift}}{\approx}0.5\,\rm{km/s}$, which is
  close to the thermally dominated regime where ice formation is most
  efficient.  However, the depletion of water vapour is only
  $\Delta=0.1{\%}$, caused by the low dust-to-gas ratio.  We
  thus find that the depletion in population II AGB stars will be
  low, even for high mass loss rate cases.

\section{Summary/Conclusions}

\label{sect:summaryconclusions}

The results of our study can be summarised as follows.

\begin{enumerate}
        \item For a given grain size, ice formation becomes more efficient as
        the mass-loss rate of the central star increases. An increase in 
        $\dot{M}$ increases the growth rate on the grains and reduces the 
        effect of sputtering, promoting ice formation.
        \item In a circumstellar envelope, small dust grains will dominate the
        ice formation process if they are present in sufficiently large 
        relative numbers. Compared to large dust particles, small particles 
        have a better abillity to grow ice, since they are less sensitive for 
        sputtering and, for a fixed $\dot{M}$, have a larger total collecting 
        area. 
        \item For small particle sizes ($a_{0}=0.01{\ }\rm{{\mu}m}$), modest 
        ice growth may already start at mass-loss rates of 
        $10^{-6}{\ }\rm{M_{\odot}/yr}$ and major ice formation only sets in 
        when the mass-loss rates are typically of the order of several times 
        $10^{-5}$ to $10^{-4}{\ }\rm{M_{\odot}/yr}$, i.e. during the superwind
        phase of an AGB star.
        \item The ice formation process can be characterised in terms of 
        $v_{\rm{drift}}$. For any given particle size, sputtering is an 
        effective mechanism to heavily suppress or completely block ice growth 
        at $v_{\rm{drift}}\geq{3{\ }\rm{km/s}}$. Between 
        $v_{\rm{drift}}{\approx}0.6{\ }\rm{km/s}$ and $3{\ }\rm{km/s}$ ice 
        formation becomes increasingly more efficient as $v_{\rm{drift}}$ 
        decreases, but no more than $20{\%}$ of the water vapour will form 
        ice. When $v_{\rm{drift}}<0.6{\ }\rm{km/s}$, more than $20{\%}$ of the
        vapour may condense. In this thermally dominated regime, it is 
        $\dot{M}$ that controls the amount of ice that growths. 
        \item Both crystalline and amorphous ice can be formed in 
        circumstellar envelopes. In the outflow, initially crystalline 
        ice will form. When the dust temperature gets below 
        ${\sim}65{\ }\rm{K}$, the molecules are immobilised on the grain, 
        preventing crystals from forming and resulting in amorphous 
        ice. A substantial fraction of the grain may be composed of amorphous
        ice ($\approx66{\%}$).
        \item For an increasing mass-loss rate, ice formation starts closer to
        the central star, the final particle size will be reached earlier, and
        the transition point where amorphous instead of crystalline ice 
        will be formed moves inwards. In the meanwhile, the depth up to which
        UV photons from the interstellar medium can penetrate into the 
        envelope, and make crystalline ice amorphous, moves outwards. An 
        increase of the initial grain size generally has modest or no effect
        on the distance dependence of ice formation. Only the depth up to 
        which UV photons from the interstellar medium can penetrate into the 
        envelope may change several orders of magnitude, increasing for larger
        particle sizes.
        \item For a star with a low optical depth envelope, the effect of the 
        UV radiation is generally important throughout most of the envelope, 
        and will most likely contribute to the formation of amorphous ice. 
        However, the effect of the ISM can be cancelled in the inner regions, 
        where the temperature exceeds $70{\ }\rm{K}$.
        \item We have improved upon an analytical equation, originally 
        derived by \citet{1985ApJ...292..487J}, that predicts the depletion 
        of water vapour from a circumstellar envelope. 
        \item Applying our model results and the modified equation of Jura 
        $\&$ Morris to OH/IR stars, RSGs and population II AGB stars, we find 
        that OH/IR stars reveal depletions up to at least $15{\%}$. 
        Population II AGB show very little depletion due to the
         low dust to 
        gas ratio of these stars. OH/IR stars and population II AGB stars
        mainly occupy the intermediate and the thermally dominated ice 
        formation regimes. Water ice will not or hardly form in the 
        circumstellar envelopes of RSGs due to the generally large drift 
        velocities of the dust in their envelopes, caused by their high 
        luminosities. 
\end{enumerate}


\bibliographystyle{aa}
\bibliography{h4027.bib}

\begin{thebibliography}{32}
\expandafter\ifx\csname natexlab\endcsname\relax\def\natexlab#1{#1}\fi

\bibitem[{Barlow(1998)}]{B_98_LWS_AGB}
Barlow, M.~J. 1998, \apss, 255, 315

\bibitem[{{Bertie} {et~al.}(1969){Bertie}, {Labb\'{e}}, \&
  {Whalley}}]{1969Bertie}
{Bertie}, J.~E., {Labb\'{e}}, H.~J., \& {Whalley}, E.~J. 1969, Chem. Phys.

\bibitem[{{Bohdansky} {et~al.}(1980){Bohdansky}, {Roth}, \&
  {Bay}}]{1980JAP....51.2861B}
{Bohdansky}, J., {Roth}, J., \& {Bay}, H.~L. 1980, Journal of Applied Physics,
  51, 2861

\bibitem[{{Bohren} \& {Huffmann}(1983)}]{1983BohrenHuffman}
{Bohren}, C. \& {Huffmann}, D.~R. 1983

\bibitem[{{Bouwman}(2001)}]{2001PhDT..........B}
{Bouwman}, J. 2001, Ph.D.~Thesis University of Amsterdam

\bibitem[{{Dominik}(1992)}]{1992ThesisCarsten}
{Dominik}, C. 1992, Thesis Technischen Universit\"{a}t Berlin

\bibitem[{{Dominik} {et~al.}(1989){Dominik}, {Sedlmayr}, \&
  {Gail}}]{1989A&A...223..227D}
{Dominik}, C., {Sedlmayr}, E., \& {Gail}, H.-P. 1989, \aap, 223, 227

\bibitem[{{Gail} \& {Sedlmayr}(1984)}]{1984A&A...132..163G}
{Gail}, H.-P. \& {Sedlmayr}, E. 1984, \aap, 132, 163

\bibitem[{{Hoogzaad}(2001)}]{2001MasterThesisSeppe}
{Hoogzaad}, S.~N. 2001, Master Thesis Vrije Universiteit (VU) Amsterdam

\bibitem[{{Hoogzaad} {et~al.}(2002){Hoogzaad}, {Molster}, {Dominik}, {Waters},
  {Barlow}, \& {de Koter}}]{2002A&A...389..547H}
{Hoogzaad}, S.~N., {Molster}, F.~J., {Dominik}, C., {et~al.} 2002, \aap, 389,
  547

\bibitem[{{Jura} \& {Morris}(1985)}]{1985ApJ...292..487J}
{Jura}, M. \& {Morris}, M. 1985, \apj, 292, 487

\bibitem[{{Kemper} {et~al.}(2002){Kemper}, {J{\" a}ger}, {Waters}, {Henning},
  {Molster}, {Barlow}, {Lim}, \& {de Koter}}]{2002Natur.415..295K}
{Kemper}, F., {J{\" a}ger}, C., {Waters}, L.~B.~F.~M., {et~al.} 2002, \nat,
  415, 295

\bibitem[{{Kessler} {et~al.}(1996){Kessler}, {Steinz}, {Anderegg}, {Clavel},
  {Drechsel}, {Estaria}, {Faelker}, {Riedinger}, {Robson}, {Taylor}, \&
  {Ximenez de Ferran}}]{1996A&A...315L..27K}
{Kessler}, M.~F., {Steinz}, J.~A., {Anderegg}, M.~E., {et~al.} 1996, \aap, 315,
  L27

\bibitem[{{Kouchi}(1987)}]{1987Natur.330..550K}
{Kouchi}, A. 1987, \nat, 330, 550

\bibitem[{{Kouchi} \& {Kuroda}(1990)}]{1990fspe.rept..193K}
{Kouchi}, A. \& {Kuroda}, T. 1990, in Formation of Stars and Planets, and the
  Evolution of the Solar System, 193--196

\bibitem[{{Kouchi} {et~al.}(1994){Kouchi}, {Yamamoto}, {Kozasa}, {Kuroda}, \&
  {Greenberg}}]{1994A&A...290.1009K}
{Kouchi}, A., {Yamamoto}, T., {Kozasa}, T., {Kuroda}, T., \& {Greenberg}, J.~M.
  1994, \aap, 290, 1009

\bibitem[{{Kwan} \& {Hill}(1977)}]{1977ApJ...215..781K}
{Kwan}, J. \& {Hill}, F. 1977, \apj, 215, 781

\bibitem[{{Lepault} {et~al.}(1983){Lepault}, {Freeman}, \&
  {Dubochet}}]{1983Lepault}
{Lepault}, J., {Freeman}, R., \& {Dubochet}, J. 1983, J. Microsc., 132, 3

\bibitem[{{Meyer} {et~al.}(1998){Meyer}, {Smith}, {Charnley}, \&
  {Pendleton}}]{1998AJ....115.2509M}
{Meyer}, A.~W., {Smith}, R.~G., {Charnley}, S.~B., \& {Pendleton}, Y.~J. 1998,
  \aj, 115, 2509

\bibitem[{{Molster} {et~al.}(2001){Molster}, {Lim}, {Sylvester}, {Waters},
  {Barlow}, {Beintema}, {Cohen}, {Cox}, \& {Schmitt}}]{2001A&A...372..165M}
{Molster}, F.~J., {Lim}, T.~L., {Sylvester}, R.~J., {et~al.} 2001, \aap, 372,
  165

\bibitem[{{Molster} {et~al.}(2002){Molster}, {Waters}, {Tielens}, \&
  {Barlow}}]{2002A&A...382..184M}
{Molster}, F.~J., {Waters}, L.~B.~F.~M., {Tielens}, A.~G.~G.~M., \& {Barlow},
  M.~J. 2002, \aap, 382, 184

\bibitem[{{Molster} {et~al.}(1999){Molster}, {Waters}, {Trams}, {Van Winckel},
  {Decin}, {van Loon}, {J{\" a}ger}, {Henning}, {K{\" a}ufl}, {de Koter}, \&
  {Bouwman}}]{1999A&A...350..163M}
{Molster}, F.~J., {Waters}, L.~B.~F.~M., {Trams}, N.~R., {et~al.} 1999, \aap,
  350, 163

\bibitem[{{Moore}(1999)}]{1999MooreC}
{Moore}, M.~H. 1999, Solid interstellar matter: The ISO revolution

\bibitem[{{Moore} \& {Hudson}(1992)}]{1992ApJ...401..353M}
{Moore}, M.~H. \& {Hudson}, R.~L. 1992, \apj, 401, 353

\bibitem[{{Olivier} {et~al.}(2001){Olivier}, {Whitelock}, \&
  {Marang}}]{2001MNRAS.326..490O}
{Olivier}, E.~A., {Whitelock}, P., \& {Marang}, F. 2001, \mnras, 326, 490

\bibitem[{{Omont} {et~al.}(1990){Omont}, {Forveille}, {Moseley}, {Glaccum},
  {Harvey}, {Likkel}, {Loewenstein}, \& {Lisse}}]{1990ApJ...355L..27O}
{Omont}, A., {Forveille}, T., {Moseley}, S.~H., {et~al.} 1990, \apjl, 355, L27

\bibitem[{{Roth} {et~al.}(1978){Roth}, {Bohdansky}, \&
  {Martinelli}}]{1978Rothetal}
{Roth}, J., {Bohdansky}, J.~J., \& {Martinelli}, P.~A. 1978, Proceedings of the
  Int. Conf. on Ion Beam Modifications of Materials

\bibitem[{{Sopka} {et~al.}(1985){Sopka}, {Hildebrand}, {Jaffe}, {Gatley},
  {Roellig}, {Werner}, {Jura}, \& {Zuckerman}}]{1985ApJ...294..242S}
{Sopka}, R.~J., {Hildebrand}, R., {Jaffe}, D.~T., {et~al.} 1985, \apj, 294, 242

\bibitem[{{Strazzulla} {et~al.}(1985){Strazzulla}, {Baratta}, \&
  {Magazzu}}]{1985AExpr...1..143S}
{Strazzulla}, G., {Baratta}, G.~A., \& {Magazzu}, A. 1985, Astronomy Express,
  1, 143

\bibitem[{{Suh}(1999)}]{1999MNRAS.304..389S}
{Suh}, K. 1999, \mnras, 304, 389

\bibitem[{{Sylvester} {et~al.}(1999){Sylvester}, {Kemper}, {Barlow}, {de Jong},
  {Waters}, {Tielens}, \& {Omont}}]{1999A&A...352..587S}
{Sylvester}, R.~J., {Kemper}, F., {Barlow}, M.~J., {et~al.} 1999, \aap, 352,
  587

\bibitem[{{Woitke} {et~al.}(1993){Woitke}, {Dominik}, \&
  {Sedlmayr}}]{1993A&A...274..451W}
{Woitke}, P., {Dominik}, C., \& {Sedlmayr}, E. 1993, \aap, 274, 451+

\end{thebibliography}

\end{document}